\documentclass{emulateapj}
\usepackage[utf8]{inputenc}
\usepackage{rotating}
\usepackage{natbib}
\usepackage{url}
\usepackage[usenames]{color}
 \usepackage{hyperref}
 \usepackage{float}
\newcommand{\bjdtdb}{\ensuremath{\rm {BJD_{TDB}}}}
\newcommand{\feh}{\ensuremath{\left[{\rm Fe}/{\rm H}\right]}}
\newcommand{\teff}{\ensuremath{T_{\rm eff}}}

\newcommand{\msun}{\ensuremath{\,M_\Sun}}
\newcommand{\rsun}{\ensuremath{\,R_\Sun}}
\newcommand{\lsun}{\ensuremath{\,L_\Sun}}

\newcommand{\rj}{\ensuremath{\,R_{\rm J}}}
\newcommand{\fave}{\langle F \rangle}
\newcommand{\fluxcgs}{10$^9$ erg s$^{-1}$ cm$^{-2}$}
\begin{document}

\title{ Validation and Initial Characterization of the  Long Period  Planet  Kepler-1654 b}
\author{C.\ A.\ Beichman}
\affil{\it NASA Exoplanet Science Institute  and Infrared Processing and Analysis Center, California Institute of Technology, Jet Propulsion Laboratory, Pasadena, CA 91125}
\email{chas@ipac.caltech.edu}
\author{H.\ A.\ C.\ Giles}
\affil{\it Observatoire Astronomique de l’Université de Genève}
\author{R.\ Akeson, D.\ Ciardi, J.\ Christiansen}
\affil{\it NASA Exoplanet Science Institute  and Infrared Processing and Analysis Center, California Institute of Technology, Jet Propulsion Laboratory, Pasadena, CA 91125}
\author{H.\ Isaacson, G.\ M.\ Marcy}
\affil{\it University of California at Berkeley, Berkeley, CA}
\author{E.\ Sinukoff}
\affil{\it Institute for Astronomy, University of Hawaii, 96822}
\author{T.\ Greene}
\affil{\it NASA Ames Research Center, Mountain View, CA, 94035 }
\author{J.\ J.\ Fortney}
\affil{\it University of California Santa Cruz, Santa Cruz, CA,95064}
\author{I. Crossfield}
\affil{\it  Massachusetts Institute of Technology, Cambridge, MA 02139 }
\author{R. Hu}
\affil{\it Jet Propulsion Laboratory, Pasadena, California Institute of Technology, 91125}
\and

\author{A.\ W.\ Howard, E.\ A.\ Petigura, H.\ A.\  Knutson}
\affil{\it California Institute of Technology, 91125}

\date{February 24, 2018}

%\begin{document}

%\maketitle

\section{Abstract}

Fewer than 20 transiting  Kepler  planets have periods longer than one year. Our early search of the Kepler light curves revealed one such system, Kepler-1654 b (originally KIC~8410697b), which  shows exactly  two transit events and whose second transit occurred only 5 days before the failure of the second of two reaction wheels brought the primary  Kepler mission to an end. A number of authors have also  examined light curves from the  Kepler mission searching for long period planets and identified this candidate.   Starting in Sept. 2014 we began an observational program of imaging, reconnaissance spectroscopy and precision radial velocity measurements which confirm with a high degree of confidence that Kepler-1654 b is a {\it bona fide} transiting planet orbiting a mature G2V star (T$_{eff}= 5580$K, [Fe/H]=-0.08) with a semi-major axis of 2.03 AU, a period of 1047.84 days and a radius of 0.82$\pm$0.02 R$_{Jup}$. Radial Velocity (RV) measurements using Keck's HIRES spectrometer obtained over 2.5 years set  a limit to the planet's mass  of $<0.5\ (3\sigma$)  M$_{Jup}$. The bulk density of the planet is similar to that of Saturn or possibly lower.  {   We assess the suitability of  temperate gas giants like  Kepler-1654b  for transit spectroscopy with the James Webb Space Telescope  since their  relatively cold  equilibrium  temperatures (T$_{pl}\sim 200$K)  make them   interesting from the standpoint of exo-planet atmospheric physics. Unfortunately, these low temperatures also make  the atmospheric  scale heights small and thus  transmission spectroscopy challenging. Finally, the long time between transits can make scheduling JWST observations difficult---as is the case with Kepler-1654b}

\section{Introduction}

The Kepler mission \citep{Borucki2010} has revolutionized our understanding of exoplanets, finding over  2,300  confirmed planets and almost 4500  candidates\footnote{As of December 2107 for Kepler with an additional 170 confirmed planets for K2, http://exoplanetarchive.ipac.caltech.edu/}\citep{Batalha2013}. These data have improved our knowledge of the constituents of the inner solar system with an inventory that includes planets ranging from less than an Earth radius (Kepler 37b) up to   1.5 Jupiter radii (Kepler 12b), and periods  ranging from less than a day (Kepler 78b) up to 1100 days, including Kepler 167 \citep{Kipping2016}  and Kepler 1647  \citep{Kostov2016}. A number of  non-transiting Kepler planets with longer periods  were identified by their radial velocity (RV) signature, e.g. Kepler 407c with a period of order 3000 days \citep{Marcy2014}. The completeness of  the Kepler catalog is poor for long period planets. These objects are hard to find {\it a priori} since the transit probability decreases with increasing semi-major axis and because fewer transits are observable in a given observing period. A smaller number of  events    reduces the total signal-to-noise-ratio (SNR) achievable by averaging  multiple transits. Most importantly, the Kepler pipeline required 3 or more potential transits before promoting a star  to become a  "Kepler Object of Interest", or KOI, worthy of further investigation. \citep{Jenkins2010}.

To avoid the Kepler pipeline's prohibition against planets with 1 or 2 transits  we analyzed Kepler light curves {\it not identified} with confirmed planets,  Kepler candidates, or KOI's. As described below, this search was rewarded with the detection of a Jupiter sized planet in a 2.87 yr (1047.836 day) period orbiting a mid G star, KIC~8410697, which we now refer to as Kepler-1654.  A more complete search for long period systems was carried out by the Planet Hunters group \citep{Wang2015} who identified a number of systems with 1 and 2 transits. In the case of Kepler-1654  they found only the first of its two transits. \citet{Foreman2016} identified seven  new  transiting systems, showing 1 or at most 2 transits, and 8 long period planets identified with known  Kepler systems having at least one shorter period planet. 

This paper describes follow-up observations of Kepler-1654 using the  W.M. Keck Observatory  have allowed us to reject a variety of alternative ("false-positive") interpretations, fully characterize the host star, and  to set an upper limit  to its mass to be less than 0.48 M$_{Jup}\, (3 \sigma)$.  $\S$\ref{lightcurves}  describes the search through the Kepler Light curves,  $\S$\ref{follow} the follow-up observations of the star, $\S$\ref{TransitProps} the  characterization of the planet, and  $\S$\ref{atmosphereJWST} investigates the prospects of studying the planet's atmosphere with JWST transit spectroscopy.

\section{Searching non-KOI Light curves\label{lightcurves}}

The data used for this investigation were drawn from Quarters 1-17 and encompassed the entire duration of the Kepler prime mission. A total of 11,232 stars were selected on the basis of their properties in the Kepler Stellar Database   \citep{Brown2011}: Kepler magnitude, Kp $<14$ mag, effective temperatures between 5500K -- 6000 K and  log g $>$3.75. These stellar values are of course only rough estimates  \citep{Huber2014} and were used only for an initial selection of likely F5-G5 dwarf stars. Data within each Quarter, $I(t)$, were normalized to near-unity using a trimmed mean signal for the entire Quarter and then searched for individual flux dips using a zero-sum Box Car filter of length $3L$ where $L$ was allowed to range in duration from 4 to 24 hours. A local trimmed average and standard deviation were evaluated within each segment with the filter output, $S(t)$.  {  A local trimmed average and standard deviation were evaluated within each segment with the filter output, $S(t)$, at a given time, $t$, having a value, }

$$   S(t)=<I(t)>_{-L/2}^{L/2} -0.5 \times (<I(t)>_{-3L/2}^{-L/2} $$
$$ \, \,  \, \,  \, \,  \, \,  \, \,  +<I(t)>_{+L/2}^{+3L/2}) \eqno{(1)}.$$

Negative going dips  with Signal to Noise Ratio (SNR)$> 20$ were output for subsequent analysis. The  noise per sample, $\sigma$, used in this calculation was derived on a  Quarter  by Quarter basis using  a robust estimate of standard deviation of all points within the Quarter\footnote{We used the ``resistant\_mean" algorithm in the GSFC IDL library,  http://idlastro.gsfc.nasa.gov/contents.html. Routines in this library were used for a number of other calculations in this work.}, $\sigma_Q$, rejecting values deviating by more than $\pm3\sigma$ from the initial mean and standard deviation. The SNR of a potential transit event was evaluated by {  dividing the depth of the event by the  noise per sample, $\sigma$, and multiplying by  $\sqrt{N_L}$ where $N_L$ is the number of samples in a segment of length $L$. }

A  list of 24 systems was  examined more closely. For most of the single transit cases, the transit duration combined with the approximate properties of the star yielded predicted  orbital periods \citep{Seager2003} much greater than duration of the Kepler mission. These systems would be impossible to confirm. In a few cases the predicted orbital periods were short compared to the mission duration, implying that the Kepler pipeline should have found and considered the object if real.

One object we identified is Kepler-1654b, orbiting a mid-G dwarf star with a Kepler magnitude of 13.42 mag, a transit depth of 0.51\%, and a period of 1047.8356 days (2.87 yr,Table~\ref{props}). \citet{Wang2015} identified this object as having only a single transit on Day 542+2454833 (BJD). By going to the very end of Q17 we were able to identify the second transit  on  Day 1590+2454833 (BJD).   \citet{Foreman2016} also found  two transits for this system.

\begin{deluxetable}{lcc}
\tablecaption{Observed Properties of Kepler-1654 \label{props} }
\tabletypesize{\scriptsize}
%\tablewidth{0.6\columnwidth} 
\tablehead{Property&Value&Comment}
\startdata
Kepler\# &1654\\
KIC \# &8410697&\\
2MASS  designation&J18484459+4426041\\
$\alpha$&18h48m44.6s&J2000\\
$\delta$&44d26m04.1s&J2000\\
Kepler Mag  &13.42              &mag\\
J            &12.28$\pm$0.021  &mag\\
H            &11.93$\pm$0.019  &mag\\
K           &11.92	$\pm$0.015  &mag\\
 WISE W1    &11.88$\pm$0.023   &mag\\
 WISE W2    & 11.92 $\pm$ 0.022  &mag\\
T$_{eff}$   &5580$\pm$ 70 K     & Keck HIRES\\
log g       & 4.19 $\pm$ 0.06   &Keck HIRES\\
$\lbrack$ Fe/H $\rbrack$         & -0.08 $\pm$ 0.06  &Keck HIRES\\
Vsini       & $<$ 2.0 km s$^{-1}$&Keck HIRES\\
Stellar Age&$>$ 5 Gyr&Keck HIRES\\\hline
\enddata 
\end{deluxetable}

\begin{figure*}
 \includegraphics[scale=0.55]{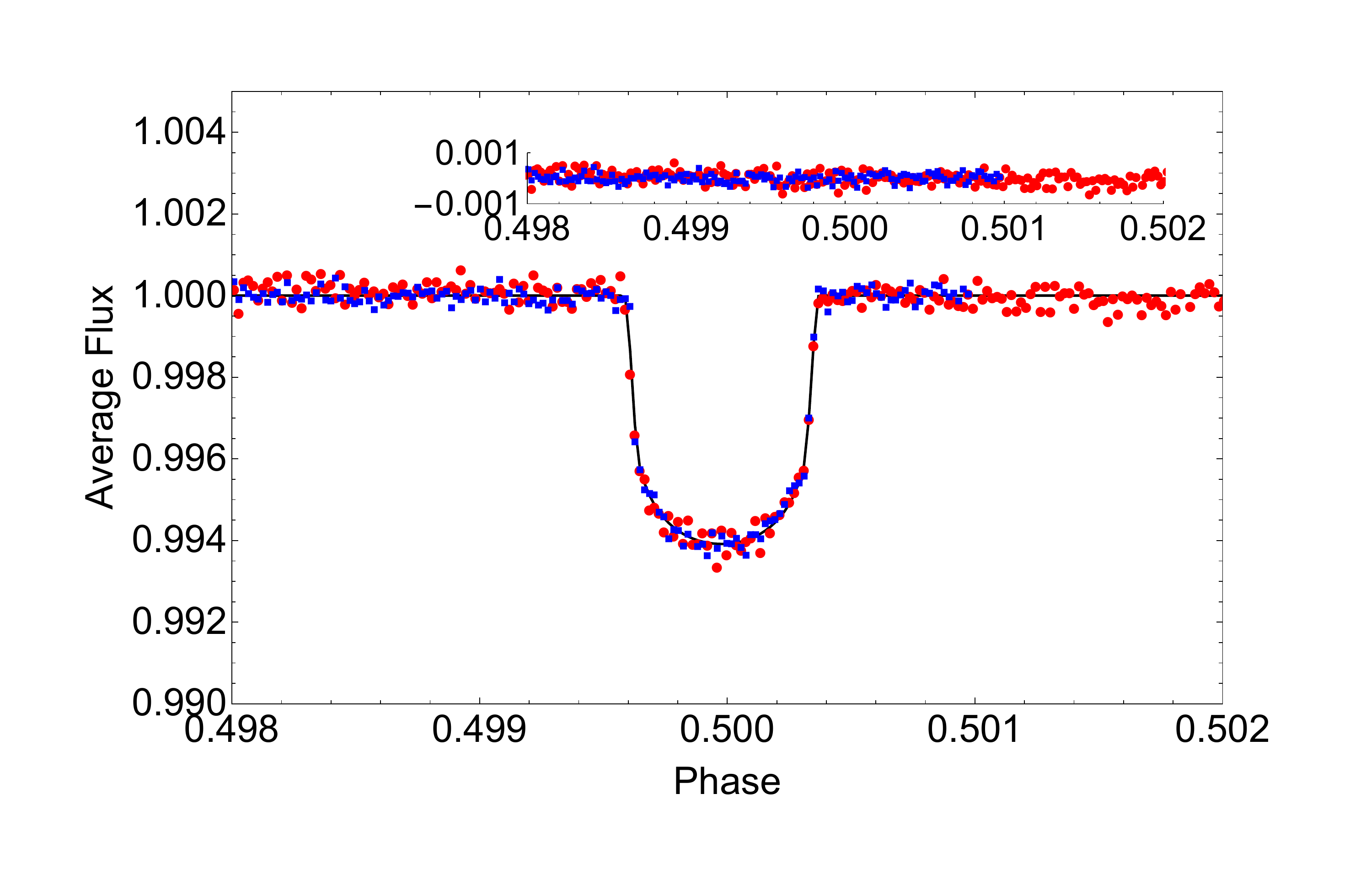}\caption{Kepler data from Quarters 6 and 17 have been normalized, detrended  with a linear (Q6, red) or second order (Q17, blue) baseline, and phased around the period of the transit. The solid line shows a fit to these data using a model based on the EXOFAST routines \citep{Eastman2013}. The inset in the upper right shows residuals with respect to the fit.    \label{TransitFit}}
  \end{figure*}

Figure~\ref{TransitFit} shows light curves from Quarters 6 and 17  which were normalized and detrended using either a linear (Q6) or 2$^{nd}$ order (Q17) baseline to remove small trends. We also examined the entire light curve looking for other transit signatures using the LombScargle tool available at the Exoplanet Archive\footnote{http://exoplanetarchive.ipac.caltech.edu/}.  No  significant periodicities indicative of shorter period planets could be identified in the periodogram.  A search through the Kepler light curve using the TERRA software (Petigura 2013) revealed no other planets in this system. This limit is approximated by a limiting depth of  $80\ ppm \times (Period/1\, day)^{0.6}$. Thus $>$ 80ppm transits with 1 day orbital periods ($\sim 1 R_\oplus$) are ruled out and 100-day planets with depths $>$ 1300 ppm ($\sim 0.35 R_\mathrm{Jup}$) are ruled out.    Nor did we find any evidence of  a transit at  half of the nominal 1047.8  day period thereby ruling out  the presence of an  eclipsing binary in an edge-on, circular orbit \citep{Santerne2013}. 

Superimposed on the light curves in Figure ~\ref{TransitFit} is  a model transit curve fitted to the data as described in $\S$\ref{exofast}. But before describing the result of the light curve analysis, we first discuss the observations used to reject false positive interpretations and to characterize more fully the star and the transiting planet.

\begin{deluxetable*}{lcc}
\tablecaption{Median values and 68\% confidence interval for EXOFAST$^*$}
\tabletypesize{\scriptsize}
%\tablewidth{0.6\columnwidth} 
\tablehead{\colhead{~~~Parameter} & \colhead{Units} & \colhead{Value}}
\startdata
\sidehead{ Stellar Parameters:}
                           ~~~$M_{*}$\dotfill &Mass (\msun)\dotfill & $1.011_{-0.052}^{+0.056}$\\
                         ~~~$R_{*}$\dotfill &Radius (\rsun)\dotfill & $1.179_{-0.023}^{+0.026}$\\
                     ~~~$L_{*}$\dotfill &Luminosity (\lsun)\dotfill & $1.23_{-0.11}^{+0.12}$\\
                         ~~~$\rho_*$\dotfill &Density (cgs)\dotfill & $0.876_{-0.033}^{+0.015}$\\
              ~~~$\log(g_*)$\dotfill &Surface gravity (cgs)\dotfill & $4.3001_{-0.012}^{+0.0099}$\\
              ~~~$\teff$\dotfill &Effective temperature (K)\dotfill & $5597_{-93}^{+95}$\\
                              ~~~$\feh$\dotfill &Metallicity\dotfill & $-0.088_{-0.095}^{+0.097}$\\
\sidehead{Planetary Parameters:}
                              ~~~$P$\dotfill &Period (days)\dotfill & $1047.8356_{-0.0019}^{+0.0018}$\\
                       ~~~$a$\dotfill &Semi-major axis (AU)\dotfill & $2.026_{-0.035}^{+0.037}$\\
                           ~~~$R_{P}$\dotfill &Radius (\rj)\dotfill & $0.819_{-0.017}^{+0.019}$\\
           ~~~$T_{eq}$\dotfill &Equilibrium Temperature (K)\dotfill & $206.0_{-3.5}^{+3.7}$\\
               ~~~$\fave$\dotfill &Incident flux (\fluxcgs)\dotfill & $0.000408_{-0.000027}^{+0.000030}$\\
       \sidehead{Primary Transit Parameters:}
                ~~~$T_C$\dotfill &Time of transit (\bjdtdb)\dotfill & $2455375.1341_{-0.0015}^{+0.0014}$\\
~~~$R_{P}/R_{*}$\dotfill &Radius of planet in stellar radii\dotfill & $0.07138_{-0.00032}^{+0.00033}$\\
     ~~~$a/R_{*}$\dotfill &Semi-major axis in stellar radii\dotfill & $370.3_{-4.7}^{+2.2}$\\
              ~~~$u_1$\dotfill &linear limb-darkening coeff\dotfill & $0.401_{-0.025}^{+0.024}$\\
           ~~~$u_2$\dotfill &quadratic limb-darkening coeff\dotfill & $0.205\pm0.034$\\
                      ~~~$i$\dotfill &Inclination (degrees)\dotfill & $89.982_{-0.017}^{+0.012}$\\
                           ~~~$b$\dotfill &Impact Parameter\dotfill & $0.114_{-0.079}^{+0.11}$\\
                         ~~~$\delta$\dotfill &Transit depth\dotfill & $0.005096_{-0.000045}^{+0.000047}$\\
                ~~~$T_{FWHM}$\dotfill &FWHM duration (days)\dotfill & $0.8933_{-0.0053}^{+0.0038}$\\
          ~~~$\tau$\dotfill &Ingress/egress duration (days)\dotfill & $0.06463_{-0.00072}^{+0.0023}$\\
                 ~~~$T_{14}$\dotfill &Total duration (days)\dotfill & $0.9580_{-0.0039}^{+0.0035}$\\
      ~~~$P_{T}$\dotfill &A priori non-grazing transit prob\dotfill & $0.002508_{-0.000015}^{+0.000032}$\\
                ~~~$P_{T,G}$\dotfill &A priori transit prob\dotfill & $0.002893_{-0.000017}^{+0.000038}$\\
                            ~~~$F_0$\dotfill &Baseline flux\dotfill & $1.0000036_{-0.0000087}^{+0.0000089}$\\
\sidehead{Secondary Eclipse Parameters:}
              ~~~$T_{S}$\dotfill &Time of eclipse (\bjdtdb)\dotfill & $2455899.05191\pm0.00091$\\
 \sidehead{From EXOFAST run with non-zero eccentricity}
      ~~~$e$\dotfill &Eccentricity\dotfill & $0.26_{-0.11}^{+0.21}$\\
    ~~~$\omega_*$\dotfill &Argument of periastron (degrees)\dotfill & $81_{-71}^{+73}$\\
\enddata
\tablecomments{$^*$Parameters derived with eccentricity forced to zero except as noted.}
\label{KIC8410697Fit}
\end{deluxetable*}

\section{Follow-up Observations of  Kepler-1654 and Kepler-1654b\label{follow}}

\subsection{Keck AO Imaging}

\begin{figure*}
 \includegraphics[scale=0.5]{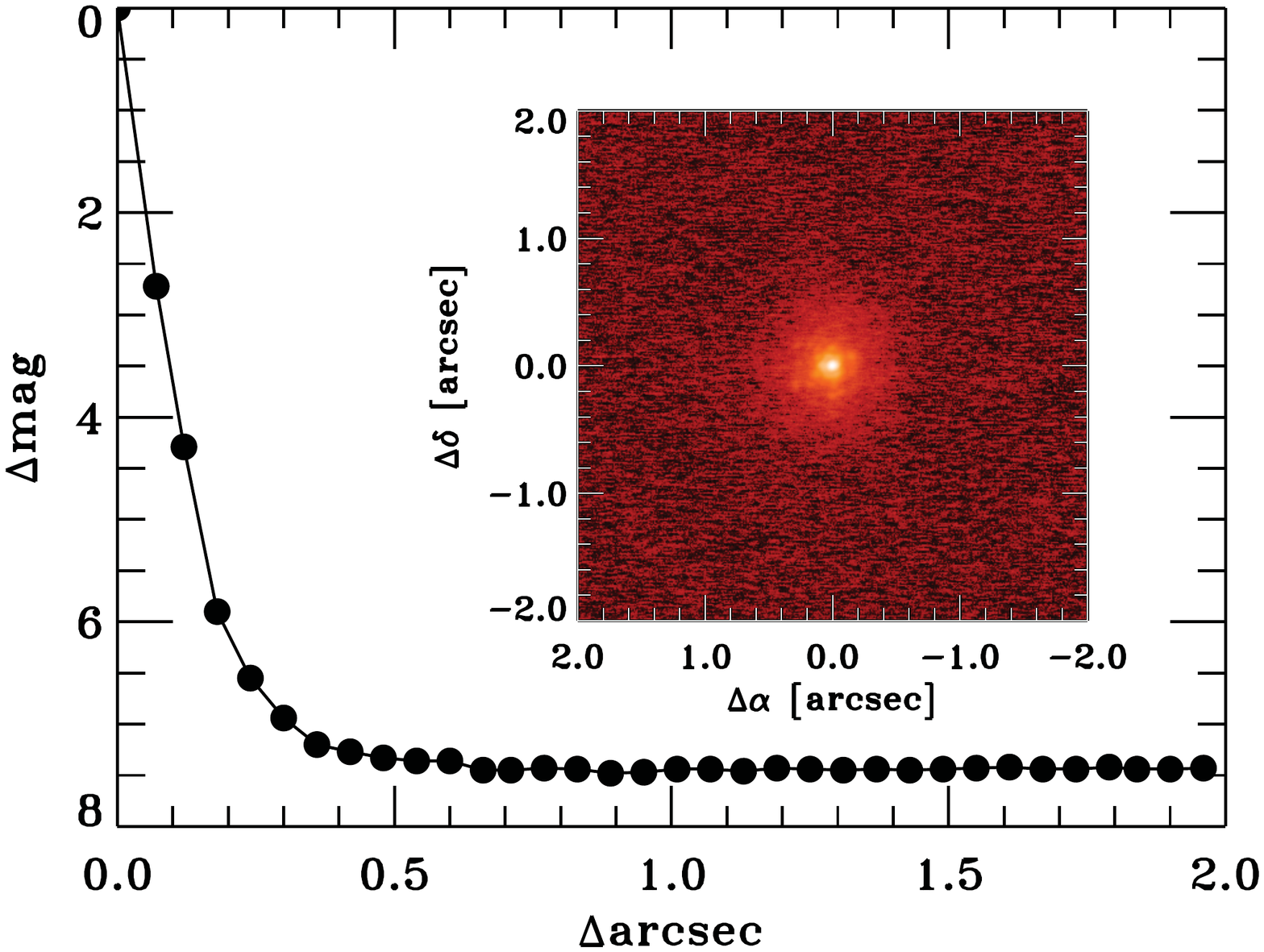}\caption{An image of Kepler-1654 obtained with the Keck II telescope in the narrow band $Br$-$\gamma$ filter shows no evidence for a companion within 4\arcsec\, of the central star.  The derived $5\sigma$ detection limits for the infrared imaging are also shown: differential magnitude as a function of angular separation from the primary star. \label{Kimage}}
 \end{figure*}

We obtained near-infrared adaptive optics images of Kepler-1654 at Keck Observatory on the night of 2015-08-21 UT (Figure~\ref{Kimage}). Observations were obtained with the 1024 $\times$ 1024 NIRC2 array and the natural guide star system; the target star was bright enough to be used as the guide star. The data were acquired using the narrow-band $Br$-$\gamma$ filter using the narrow camera field of view with a pixel scale of 9.942 mas/pixel. The $Br$-$\gamma$ filter has a narrower bandwidth (2.13–2.18 $\micron$), but a similar central wavelength (2.15 $\micron$) compared the Ks filter (1.95-2.34 $\micron$; 2.15 $\micron$) and allows for longer integration times before saturation. A 3-point dither pattern was utilized to avoid the noisier lower left quadrant of the NIRC2 array. The 3-point dither pattern was observed three times with 2 coadds and a 30 second integration time per coadd for a total on-source exposure time of $3\times 3 \times 2 \times30s = 540 s.$

The target star was measured with a resolution of 0.059\arcsec\ (FWHM). No other stars were detected within the 10\arcsec\ field of view of the camera. In the $Br$-$\gamma$ filter, the data are sensitive to stars that have K-band contrast of $\Delta$K = 4.3 mag at a separation of 0.1\arcsec\ and $\Delta$K = 7.49 at 0.5\arcsec\ from the central star. We estimate the sensitivities by injecting fake sources with a signal-to-noise ratio of 5 into the final combined images at distances of N $\times$ FWHM from the central source, where N is an integer. The 5$\sigma$ sensitivities, as a function of radius from the star, are also shown in Figure~\ref{Kimage}.

There is a star 7\arcsec\ northwest of Kepler-1654 that was outside the field of view of the NIRC2 observations.  However, this star is clearly resolved in 2MASS and is a separate star in the Kepler Input Catalog (KIC 8410692).  The KIC photometry of KIC 8410692  (KepMag=17.64 mag) indicates that the star has an effective temperature and a surface gravity of $T_{eff}=6111$K and $\log{g} = 4.35$, making the star a main sequence F dwarf at a distance of about 4 kpc, and, thus, not a bound companion to Kepler-1654.  The Kepler photometric aperture is oriented such that the background star is not included in the aperture in quarters 6 and 17 when the transits were observed, and the photocentric position remains centered on the Kepler-1654 during the transit, indicating that the transit occurs around the Kepler-1654 and not the background star. Further, at $50\times$ fainter than Kepler-1654 the photometric blending of the background star (if the entire stellar profile were inside the photometric aperture) would only dilute the observed transit, and, hence, the derived planetary radius, by $<1\%$ \citep{Ciardi2015}.

\subsection{Keck HIRES Spectroscopy}

\begin{figure*}
 \includegraphics[scale=0.75]{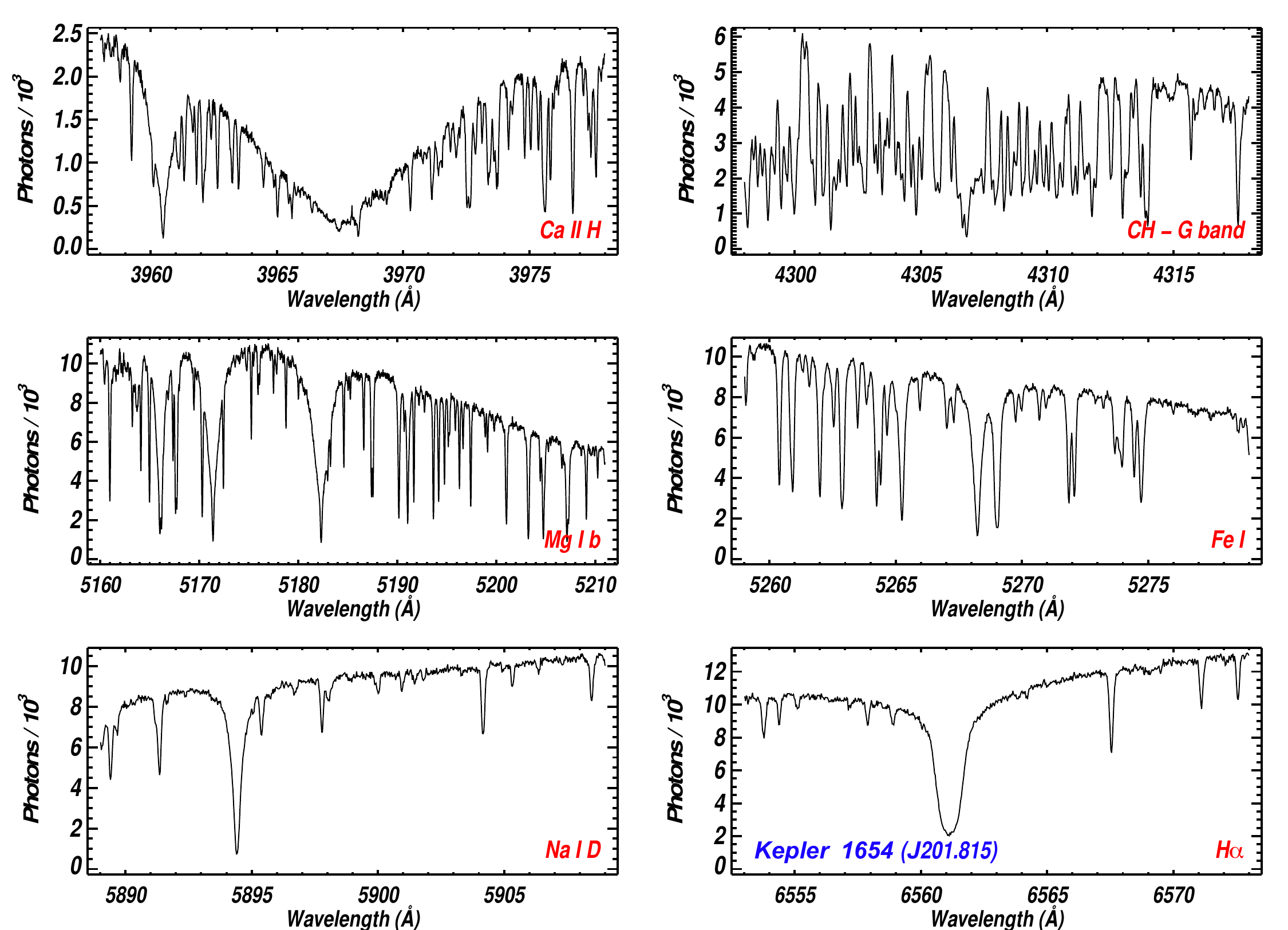}\caption{Spectra from the HIRES instrument on the Keck telescope. The top left   panel shows a portion of the spectrum near the Ca  H\&K lines and the lack of an emission reversal implies a stellar age greater than $\sim$5 Gyr.   The  middle left and bottom right  panels show lines  near the Mg b triplet and  H-$\alpha$ which  look normal for a mid-G star with narrow lines.  \label{spectra}}
 \end{figure*}

We obtained   spectra of Kepler-1654 using the HIRES instrument \citep{Vogt94} at the W.\ M.\ Keck Observatory. Observations and data reduction followed the usual methods of the California Planet Search \citep[CPS;][]{Howard2010}.  A spectrum   obtained with a 15 minute exposure on 2014/9/14 without the iodine cell was used for spectral typing (Figure~\ref{spectra}).   The  spectral synthesis modeling program  ``SpecMatch" \citep{Petigura2015} has been calibrated with asteroseismology stars and  yielded values of T$_\mathrm{eff}$, log g, and [Fe/H] with formal uncertainties of 70 K, 0.06 dex, and 0.06 dex, respectively (Table ~\ref{props}). These parameters show the star is a slowly rotating, G5 main sequence star, perhaps beginning to evolve off the main sequence.  The Ca H\&K lines  show  no emission reversal implying a stellar age greater than $\sim$5 Gyr. An analysis  looking for  secondary spectra in the HIRES spectrum of Kepler-1654 found no companions brighter than 1\% of the primary \citep{Kolbl2015}. These stellar values are similar to those cited in \citep{Foreman2016}: our spectroscopically derived values of  (T$_{eff}$,R$_*$)=({ 5580$\pm$70 K,  1.18$\pm$0.03 R$_\odot$) } vs. (5918$\pm$160 K,$1.0^{+0.35}_{-0.16}$ R$_\odot$) for Foreman-Mackey's values. We adopt our stellar values in this analysis (Table~\ref{props} and \ref{KIC8410697Fit}).

We collected 18 RV measurements  between 2014/09/07 and 2017/3/30.  An iodine cell was used for each observation as a wavelength calibrator and point spread function (PSF) reference.  Each spectrum spanned wavelengths from 3600–-8000 $\AA$ with a spectral resolution of R=60,000 and typical SNR per pixel of 100–200. The ``C2'' decker ($0\farcs87$ $\times$ 14\arcsec\ slit) provided spectral resolution $R \sim$ 55,000 and allowed for the sky background to be measured and subtracted.  An exposure meter was used to automatically terminate exposures after reaching a target signal-to-noise ratio (SNR) per pixel at 550 nm.  The standard CPS Doppler pipeline was used to measure RVs \citep{Marcy1992, Howard2009}.   RV measurements are listed in Table~\ref{RVdata}.  These values are consistent with the transit interpretation, showing variations of $<$10 m s$^{-1}$,  ruling out definitively the false alarm possibility of an eclipsing binary which would show RV variations of  a few km s$^{-1}$ on this timescale. 

 \begin{deluxetable}{lcc}
\tablecaption{Keck HIRES Data for  Kepler-1654}
\tabletypesize{\scriptsize}
\tablehead{\colhead{JD Date} & \colhead{Velocity (m s$^{-1}$)} & \colhead{$\sigma$Vel (m s$^{-1}$)}}
\startdata
2456907.899441 & 4.61 & 3.42 \\
2457061.166912 & 13.98 & 5.52 \\
2457062.168017 & -1.34 & 5.19 \\
2457151.051283 & -7.03 & 3.13 \\
2457180.021422 & -0.33 & 3.78 \\
2457201.026295 & -16.67 & 3.90 \\
2457203.095994 & 11.25 & 4.68 \\
2457211.936684 & 9.66 & 3.31 \\
2457229.053345 & -2.74 & 3.63 \\
2457326.714999 & -3.32 & 3.21 \\
2457353.692574 & 8.66 & 3.21 \\
2457354.729928 & 7.67 & 4.20 \\
2457478.135425 & -2.43 & 3.72 \\
2457521.001943 & -8.54 & 3.06 \\
2457601.046180 & -7.80 & 5.64 \\
2457620.898209 & 7.04 & 3.33 \\
2457672.824144 & 2.84 & 3.76 \\
2457830.140368 & -19.56 & 3.38 \\
\enddata
\label{RVdata}
\end{deluxetable}

\section{Analysis of the Transit and RV Observations\label{TransitProps}}

\subsection{Properties of the Transiting Planet Kepler-1654b\label{exofast}}

First it is important to to confirm that this system truly represents a giant transiting planet. We used the VESPA tool to estimate \citep{Morton2012,Morton2015} {  the probability that this signal represents an astrophysical false positive}.  As inputs, we used the light curve shown in Fig.~1, the stellar parameters listed in Table~1 along with {\it gri} photometry from APASS, the NIRC2 contrast curve described in Sec.~3.2, the Keck/HIRES limit on secondary spectra of $\Delta$mag$<$5, and an upper limit on any secondary eclipse of $2\times10^{-4}$.  The most likely false positive configuration is that of a blended eclipsing binary, but this scenario is roughly 20,000 times less likely than the planetary scenario. The resulting false positive probability is $6.2\times10^{-5}$, more than sufficient to validate Kepler-1654 as a transiting planet. {   \citet{Foreman2016} cited a false alarm rate due to eclipsing binaries of 0.05 based on statistical estimates of  the contamination by background objects. Our much higher confidence level is due  the follow-up observations which gave direct and sensitive  limits on stellar companions as well as taking advantage of  improved stellar  parameters. {\it  It is on this basis that we suggest  Kepler-1654b (n\'ee KIC~8410697b) should be regarded as a fully confirmed Kepler object.}}

To determine the properties of the transiting  companion we used the EXOFAST transit analysis routine \citep{Eastman2013} using stellar properties derived from the Keck data as priors plus the transit light curves  as input\footnote{We used the implementation of  EXOFAST available at the NASA Exoplanet Science Institute: https://exoplanetarchive.ipac.caltech.edu/cgi-bin/ExoFAST/nph-exofast.}. We ran   EXOFAST in its full MCMC mode with the  eccentricity set  to zero  with the presented in Table~\ref{KIC8410697Fit} and shown in Figure~\ref{TransitFit}. With 715 data points in the two observed transits the $\chi^2$ of the fit was 692.5 and the rms of the residuals was  0.00024 as shown in the figure. The various fitted parameters are astrophysically reasonable. For example, the derived limb-darkening coefficients of 0.40$\pm$0.02 and 0.20$\pm$0.03 are consistent with values appropriate to the stellar properties \citep{Claret2011}. The EXOFAST fit shows the planet to be a Jupiter sized object, 0.82 R$_\mathrm{Jup}$, in a 2.03 AU orbit. At this location the equilibrium temperature of the planet is 206 K assuming an albedo of zero.

Finally, we conducted a separate fit to the transit light curve using the BATMAN software package \citep{Kreidberg2015}.  All light curve parameters from this analysis agree with those in Table~\ref{KIC8410697Fit} to within 1$\sigma$.  Using our posterior distributions, we computed the posterior of the stellar density under the assumption of a circular orbit \citep{Seager2003}.  With the stellar density derived from our spectroscopic analysis, we then used the density posterior to investigate the photoeccentric effect \citep{Dawson2012}.  The photoeccentric effect allows a direct and independent constraint on a transiting planet's orbital eccentricity through the observable impact of any nonzero orbital eccentricity on the transit light curve.  Our analysis shows a  preference for nonzero orbital eccentricity: we find $e=0.3^{+0.3}_{-0.1}$,  {   consistent with the weakly non-zero estimate from EXOFAST when run with eccentricity as a free parameter,  $0.26_{-0.11}^{+0.21}$  (Table~\ref{KIC8410697Fit}). } The BATMAN analysis sets a lower limit on the eccentricity of $e>0.06$ at 99.7\% confidence.  Thus, like most other giant, long-period exoplanets known from radial velocity surveys, Kepler-1654b may also have  an orbital eccentricity greater than that of Jupiter and Saturn. Finally, we note our derived planet values are consistent with those derived by \citet{Foreman2016}, e.g.  $R_p=0.82\pm0.06$ vs. 0.70$\pm0.1$ for \citep{Foreman2016}.

\subsection{Precision RV: Constraining Kepler-1654b}

\begin{figure*}[!h]
\centering
\includegraphics[width=6.5in]{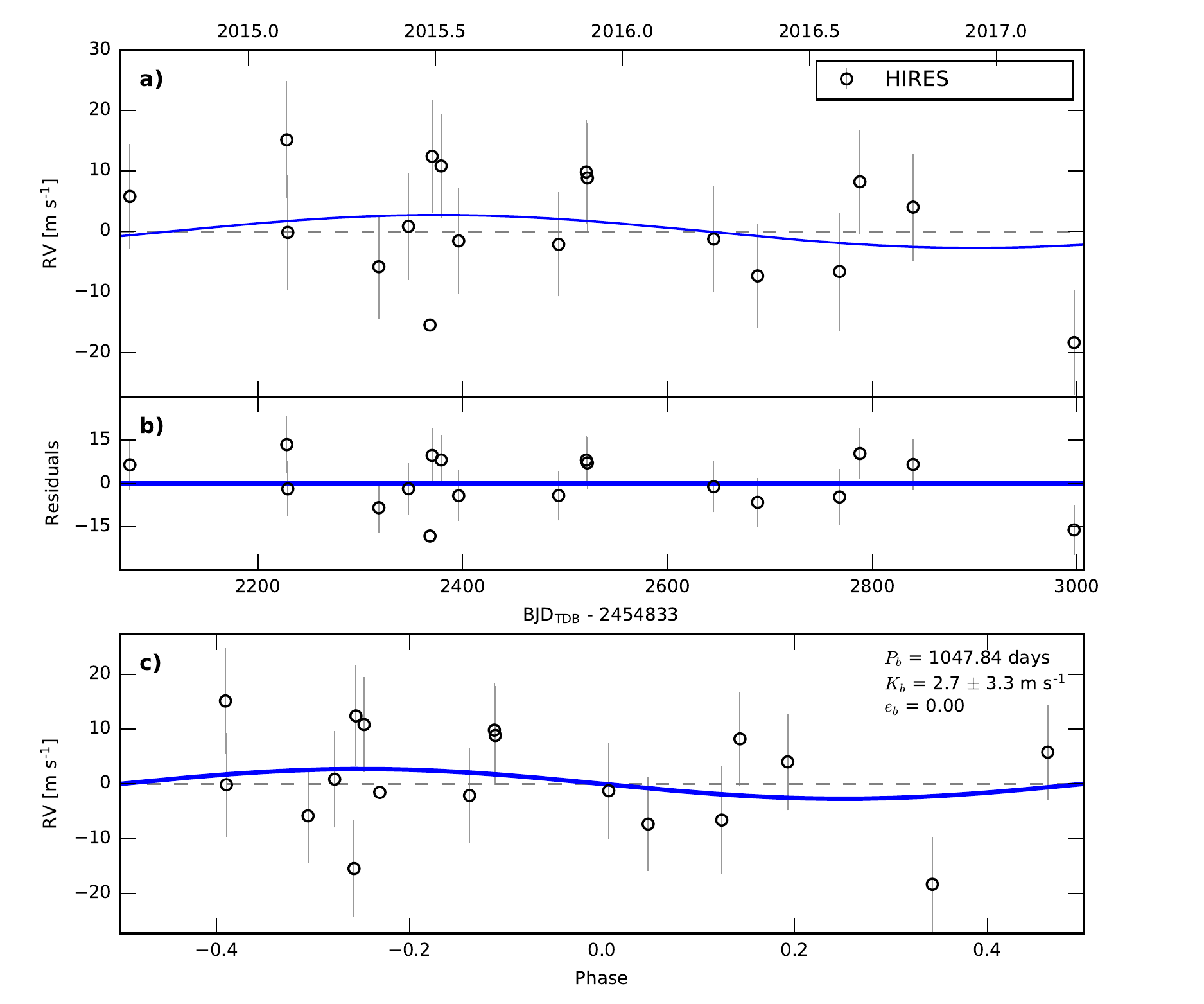}
\caption{
Best-fit 1-planet Keplerian orbital model for Kepler-1654b. The maximum likelihood model is plotted while the orbital parameters listed in Table \ref{tab:params} are the median values of the posterior distributions.
The thin blue line is the best fit 1-planet model. We add in quadrature the RV jitter term(s) listed in Table \ref{tab:params}
with the measurement uncertainties for all RVs. {\bf b)} Residuals to the best fit 1-planet model.
{\bf c)} RVs phase-folded to the ephemeris of planet b. The small point colors and symbols are the same as in panel {\bf a}. The phase-folded model for planet b is shown as the blue line. \label{fig:multiplot2}
}
\end{figure*}

\begin{figure*}[!tbp]
  \centering
  \begin{minipage}[b]{0.45\textwidth}
    \includegraphics[width=\textwidth]{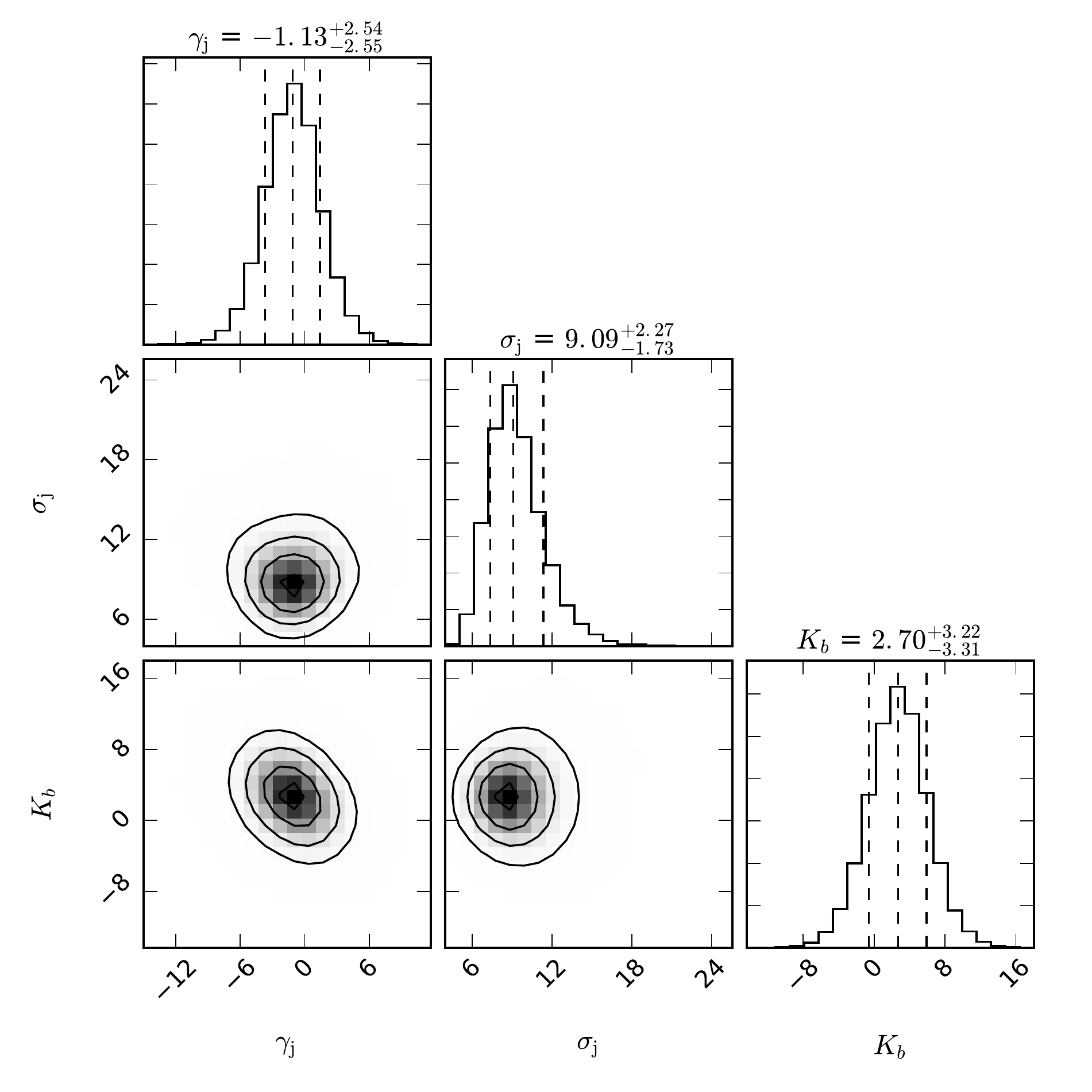}
    \caption{Posterior distributions for all free parameters.\label{freepost} }
  \end{minipage}
  \hfill
  \begin{minipage}[b]{0.45\textwidth}
    \includegraphics[width=\textwidth]{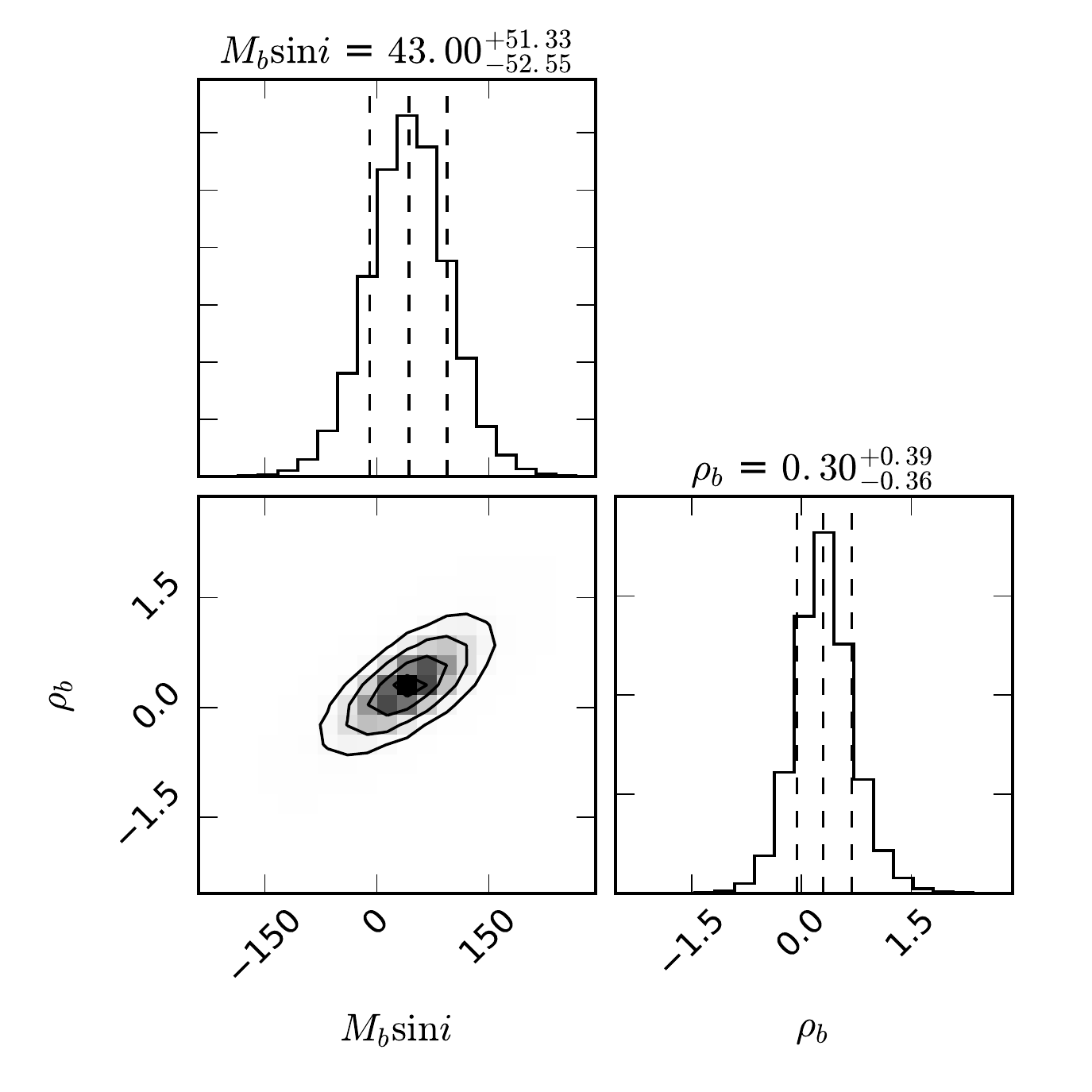}
    \caption{Posterior distributions for all derived parameters. \label{allpost}}
  \end{minipage}
\end{figure*}

 Although our RV measurements have helped to confirm the planetary nature of Kepler-1654b, our goal of  determining the  mass of the transiting planet has not yet been achieved.   We analyzed the 18 HIRES RV measurements (Table~\ref{RVdata}), which span 2.5 years, using the open source Python package \texttt{RadVel} \citep{Fulton2018}. We adopt an RV model consisting of a single Keplerian orbit, with orbital period and phase fixed at the known values and assuming an eccentricity of zero.  The model includes a constant RV offset, $\gamma$, and a ``jitter'' term $\sigma$ representing astrophysical and instrumental noise. The  MCMC analysis  (Tables~\ref{tab:comp} and ~\ref{tab:params})  yield  an  estimate of  the semi-amplitude $K_b=2.7^{+3.2}_{-3.3}$ m s$^{-1}$  which corresponds to  43$\pm$52 M$_\oplus$ (0.13$\pm$0.16 M$_{Jup}$), or a 3-$\sigma$ upper limit of $<$156  M$_\oplus$ ($<$0.49 M$_{Jup}$). Figure~\ref{fig:multiplot2} shows the RV data plotted along with the best fit model while Figures ~\ref{freepost} and ~\ref{allpost} show the posterior distributions of the model parameters.    A 0-planet model is favored on the basis of the Bayesian Information Criterion (Table~\ref{tab:comp}), consistent with a non-detection. 
 
 What level of signal might we   expect to find on the basis of a planet of radius 0.82 R$_{Jup}$? The radius-mass data shown in Figure~3 of  \citet{Howard2013a} suggest  that with a radius of 9.2 R$_\oplus$,  Kepler-1654b should  have a mass in the range of 50--100 M$_\oplus$.  \citet{Wolfgang2016} give a number of radius-mass relationships for planets with $R<4R_\oplus$ (somewhat smaller than Kepler-1654b)  and  their Method-1 yields a mass estimate of 58  M$_\oplus$ which falls within the \citet{Howard2013a}  range.  These masses correspond to RV semi-amplitudes of 3--6 m s$^{-1}$  which our  RV data only begin  to  constrain.

%\begin{figure*}[!h]
%\centering
%\includegraphics[width=6.5in]{figure5.pdf}
%\caption{Posterior distributions for all free parameters. \label{freepost}}
%\end{figure*}

%\begin{figure*}[!h]
%\centering
%\includegraphics[width=6.5in]{figure6.pdf}
%\caption{Posterior distributions for all derived parameters. \label{allpost}}
%\end{figure*}

\begin{figure*}[!h]
\centering
\includegraphics[width=6.5in]{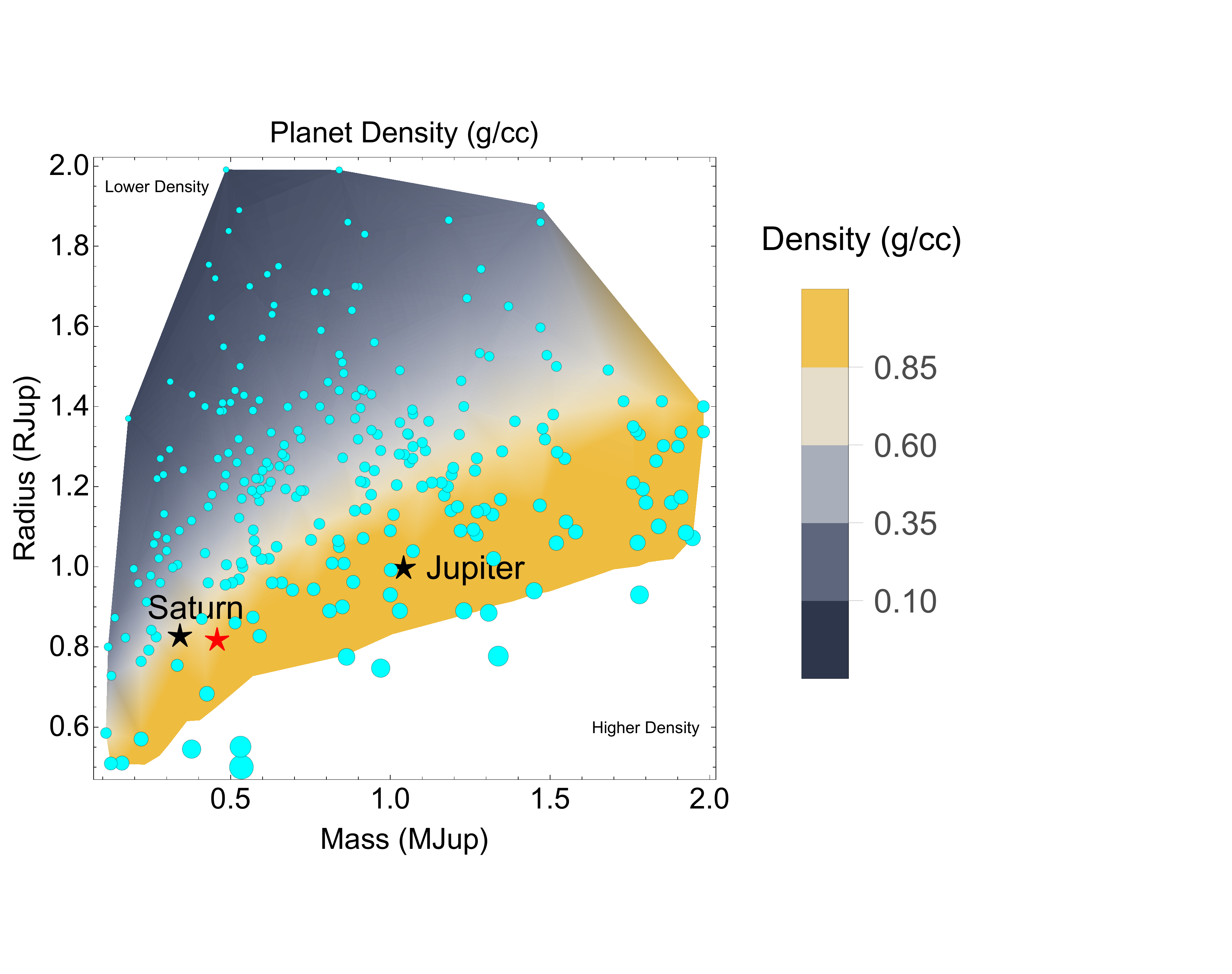}
\caption{The distribution of bulk density in g cm$^{-3}$ for planets with radii in the range of 0.5-2 R$_{Jup}$ based on data for over 250 transiting planets with well determined mass and radius measurements (cyan points). The color scale shows bulk densities from 0.1 to $>$1 g cm$^{-3}$ and shows the fall-off in bulk density for more massive planets \citep{Howard2013}.
The positions of Saturn, Jupiter and the upper limit to Kepler-1654b (red star) are indicated. The point size is proportional to the planet's density. \label{fig:massraddens2}}
\end{figure*}

The corresponding upper limit to the bulk density is $<$1.2 g cm$^{-3}$. As shown in Figure~\ref{fig:massraddens2}, the limit to Kepler-1654b's density sits close to Saturn's in the Mass-Radius-Density parameter space. Our RV observations rule out the most massive planets, but are  consistent with the distribution of planetary densities in this radius range. Continuing RV observations will eventually yield a  mass for the transiting system.

We can use our data RV to explore the upper limit to the  mass of the any interior  planet. With a 1-$\sigma$ RMS residual of 9.1  m s$^{-1}$ (Table~\ref{tab:comp}) and 18 observations we can set a 3-$\sigma$ upper limit to the RV semi-amplitude any interior planet (transiting or not) of $K=3\times9.1/\sqrt{18}=6.4$ m s$^{-1}$ for a low inclination planet where K is given by: 

$$ K=\frac{28.4\, {\rm m\, s^{-1}}} {\sqrt{1-e^2}} M_{pl}\, {\rm sin(i)}  \, M_*^{-2/3}\, P^{-1/3} \eqno{(2)}$$

\noindent with the planet mass with the planet mass in Jupiter units, the stellar mass in solar units and the period in years \citep{Lovis2010}. Assuming $sin(i)=1$ for a system with at least one transiting planet, a stellar mass of 1 M$_\odot$, and $e=0$, the HIRES observations set a mass limit for any additional planet  of $ M_{pl}<0.23  P_{yr}^{1/3}$ M$_{Jup}$.

\begin{deluxetable}{lrr}
\tablecaption{Model Comparison}
\tablehead{\colhead{Statistic} & \colhead{0 planets} & \colhead{{1 planet}}}
\startdata

$N_{\rm data}$ (number of measurements)  & 18 & 18\\
$N_{\rm free}$ (number of free parameters)  & 2 & 3\\
RMS (RMS of residuals in m s$^{-1}$)  & 9.12 & 8.88\\
$\chi^{2}$ (assuming no jitter)  & 69.91 & 67.22\\
$\chi^{2}_{\nu}$ (assuming no jitter)  & 4.37 & 4.48\\
$\ln{\mathcal{L}}$ (natural log of the likelihood)  & -65.29 & -64.87\\
BIC (Bayesian information criterion)  & 135.48 & 135.62\\

\enddata
\label{tab:comp}
\end{deluxetable}

\begin{deluxetable}{lrr}
\tablecaption{MCMC Posteriors}
\tablehead{\colhead{Parameter} & \colhead{Value} & \colhead{Units}}
\startdata
\sidehead{\bf{Modified MCMC Step Parameters}}
$\sqrt{e}\cos{\omega}_{b}$ & $\equiv$ 0.0  & \\
$\sqrt{e}\sin{\omega}_{b}$ & $\equiv$ 0.0  & \\
\hline
\sidehead{\bf{Orbital Parameters}}
$P_{b}$ & $\equiv$ 1047.8363  & days\\
$T\rm{conj}_{b}$ & $\equiv$ 2455375.133  & JD\\
$e_{b}$ & $\equiv$ 0.0  & \\
$\omega_{b}$ & $\equiv$ 0.0  & degrees\\
$K_{b}$ & 2.7 $^{+3.2}_{-3.3}$ & m s$^{-1}$\\
\hline
\sidehead{\bf{Other Parameters}}
$\gamma$ (RV offset)& -1.1 $\pm 2.6$ & m s$^{-1}$\\
%$\dot{\gamma}$ & $\equiv$ 0.0  & m s$^{-1}$ day$^{-1}$\\
%$\ddot{\gamma}$ & $\equiv$ 0.0  & m s$^{-1}$ day$^{-2}$\\
$\sigma$ (jitter) & 9.1 $^{+2.3}_{-1.7}$ & $\rm m\ s^{-1}$\\
\enddata
%\tablecomments{62500 links saved;
%Reference\  epoch for $\gamma$, $\dot{\gamma}$, $\ddot{\gamma}$: 2457369.0}
\label{tab:params}
\end{deluxetable}

Of primary importance will be to follow the Kepler-1654 system with additional RV monitoring to determine the planet's mass. New imaging and RV observations are planned to investigate the new long-period systems found by \citet{Foreman2016}.

\section{Characterizing the Atmosphere of  Temperate Gas Giants\label{atmosphereJWST}}

Kepler-1654b is representative  of the few  temperate, transiting gas-giants available for atmospheric characterization.  We investigated whether this system  might be promising for spectroscopy with the Hubble (HST) and James Webb Space Telescopes \citep[JWST;][]{Beichman2014}.  Kepler-1654b and others like it as described in \citet{Wang2015} and   \citet{Kipping2016} (Table~\ref{LongP})  will be the  coolest gas planets ($\sim200$ K) for which we will be able to probe  atmospheric composition and physical characteristics. Comparisons to planets in our own Solar System will be particularly valuable. 
 
 %when it transits again in 2021 (Table~\ref{future}).

\begin{figure*}
\centering
\includegraphics[width=6.5in]{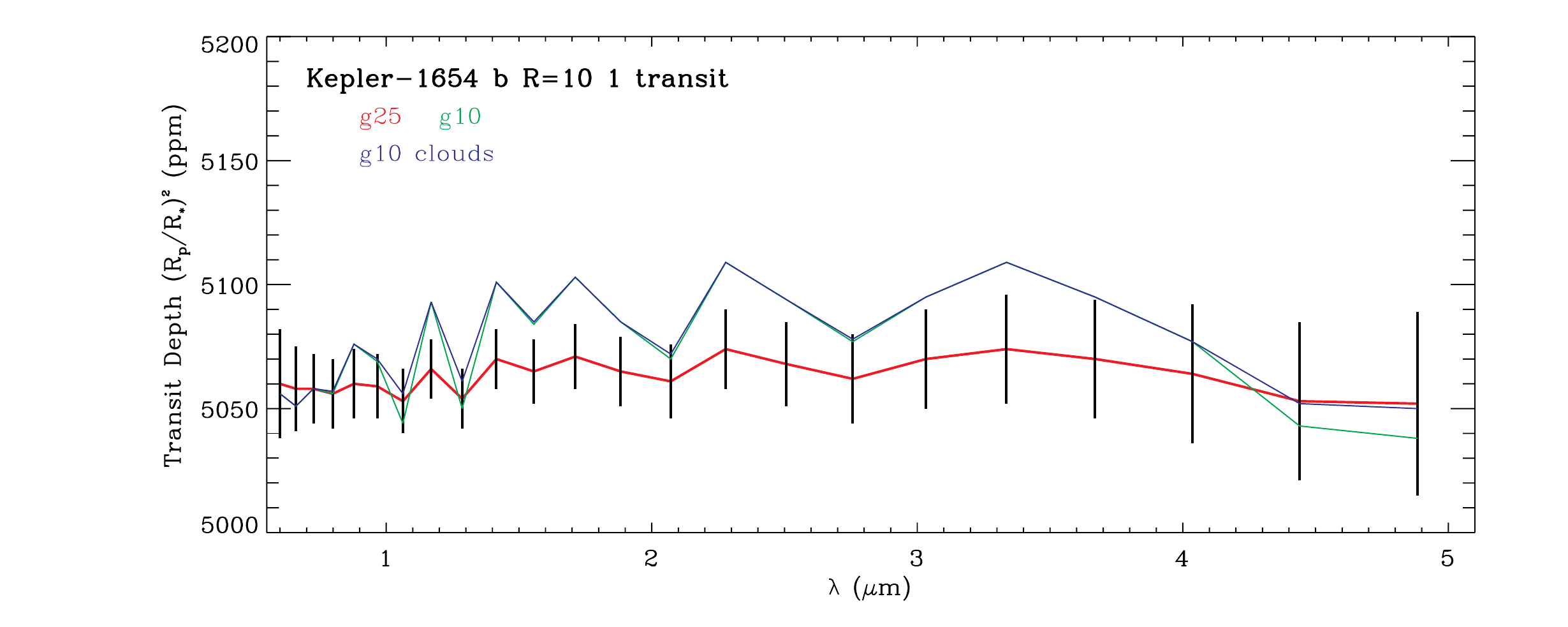}\\
\includegraphics[width=6.5in]{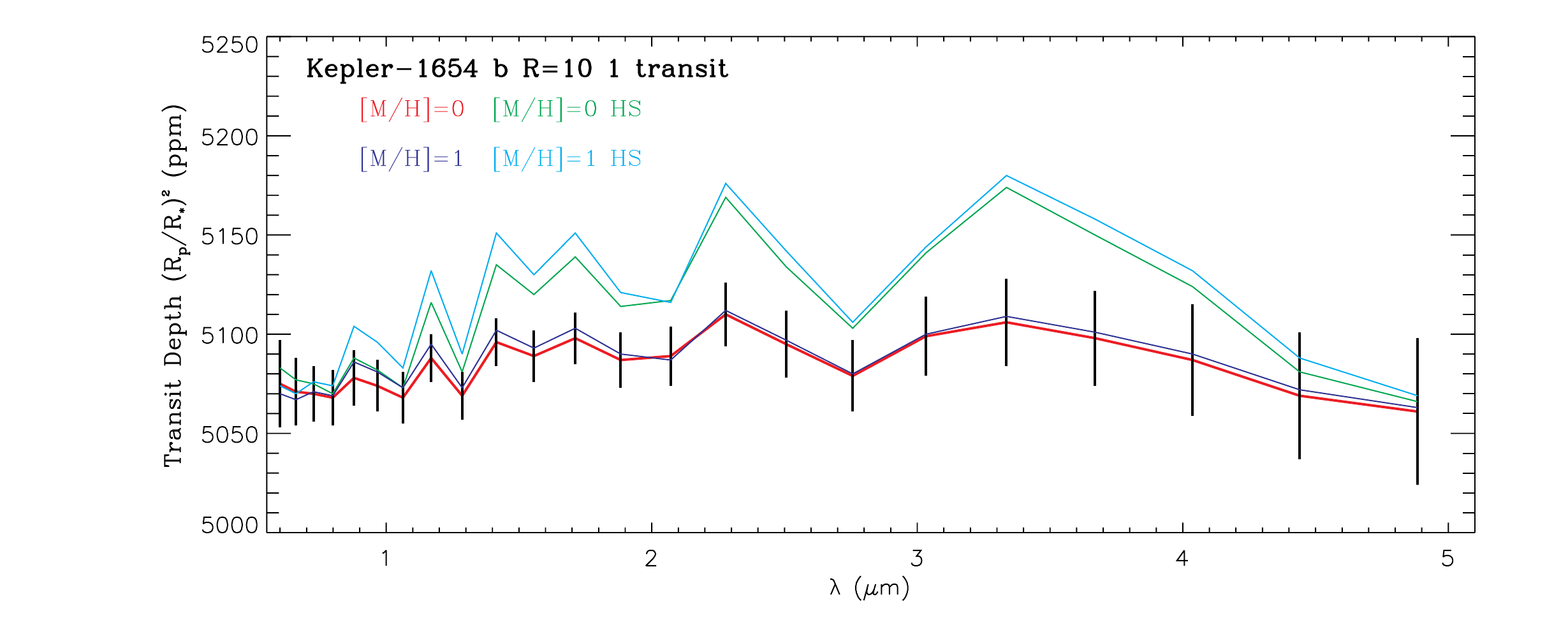}
\caption{top) Simulated JWST NIRSpec prism spectrum of Kepler-1654b with uncertainties computed as described in \citet{Greene2016}. Model spectra have been binned to $R=10$ and are shown as solid colored curves. 1 $\sigma$ uncertainties were computed for a single 18.4 hr transit at $R=10$ and are shown as error bars. Three atmospheric models with g=10 m s$^{-2}$ with and without clouds and g=25 m s$^{-2}$ without clouds appear in the top panel. The bottom panel shows $R = 10$ models and uncertainties for g=25 m s$^{-2}$ atmospheres with and without a heated stratosphere (HS) and two levels of metallicity, Solar and 10 $\times$ Solar.\label{JWST}}
\end{figure*}

Kepler-1654b is  cold for a transiting planet. The strength of absorption features in transmission spectra are proportional to a planet's atmospheric scale height, and that scale height is proportional to atmospheric temperature. Therefore the low temperature of the planet produces a small amplitude transmission spectrum. This plus the relative faintness of  Kepler-16547  itself limits the signal-to-noise of  its transmission spectrum. On the plus side, the long  duration of these events  enhances the  sensitivity for  measurements of  trace atomic and molecular species in the 1-5 $\mu$m band.  Sample spectra in the visible and near-IR for  Kepler-1654b are shown in Figure~\ref{JWST}.

\begin{deluxetable*}{ccc}
\tablecaption{Predicted Epochs of Future Transits for Kepler-1654b (UT)\label{future}}
\tabletypesize{\scriptsize}
\tablehead{\colhead{Orbit} & {Transit Midpoint (BJD)} & \colhead{Transit Midpoint (UT)} }
\startdata
0&2,455,375.1341$\pm$0.0014&2010-Jun-27 15:13:06$\pm$120 (sec)\\
1&2,456,422.9697$\pm$0.0024&2013-May 10 11:16:22$\pm$200 (sec)\\
2&2,457,470.8053$\pm$0.0040&2016-Mar 23 07:19:38$\pm$350 (sec)\\
3&2,458,518.6409$\pm$0.0059&2019-Feb-4 03:22:54$\pm$510 (sec)\\
  4& 2,459,566.4765$\pm$0.0077&2021-Dec-17 23:26:10$\pm$670 (sec)\\
{\bf 5}&{\bf 2,460,614.3121$\pm$0.0096}&{\bf 2024-Oct-30  19:29:25$\pm$830 (sec)}\\
6&2,461,662.1477$\pm$0.0115&2027-Sep-13 15:32:41.3$\pm$990 (sec)\\
7&2,462,709.9833$\pm$0.0134&2030-Jul-27 11:35:57.1$\pm$1,160 (sec)\\
8&2,463,757.8189$\pm$0.0153&2033-Jun-9 07:39:13.0$\pm$1,320 (sec)\\
\enddata
\tablecomments{These predicted transit midpoints assume no offsets  due to interactions with other bodies in the system (Transit Timing Variations, TTVs). The bold entry for 2024/10/30 is nominally the first one observable by JWST and occurs at the edge of the JWST observablity window.}
\end{deluxetable*}

We have simulated $JWST$ NIRSpec prism spectra for a single transit of this system and show the results in Figure~\ref{JWST}. These spectra were computed using the method described in \citet{Greene2016} and use Nextgen \citep{HAB99} stellar models with the T$_{eff}$ and log $g$ of Kepler-1654 (Table~1) and our atmospheric transmission models of Kepler-1654b. We model the atmosphere by solving radiative and chemical equilibrium, and also include condensation of water when supersaturation is reached. Three atmospheric models with g=10 m s$^{-2}$ with and without clouds and g=25 m s$^{-2}$ without clouds are shown  in the top panel of  Figure~\ref{JWST}a.  We computed signals in photo-electrons using the apparent stellar magnitude of  Kepler-1654 in the relevant bands,  18 hours integration time on transit, an additional 18 hours on the star, the 25-m$^2$ collecting area of $JWST$, and NIRSPEC prism resolving power and system transmission values kindly provided by the NIRSpec team (S. Birkmann, private communication). The resultant 1 $\sigma$ noise values are on the order of 15 ppm when binned to $R = 10$, lower than the best values achieved with $HST$ WFC3 G141 observations \citep[e.g.,][]{KBD14a}. It is uncertain whether $JWST$ NIRSpec or other $JWST$ instruments will achieve such low noise levels, so Figure~\ref{JWST} represents the best performance that $JWST$ is likely to achieve on a single transit observation of this system.

A second  scenario for the planet's atmosphere includes  enhanced transmission spectral features from a  heated stratosphere (Figure~\ref{JWST}b). Our models predict that the transmission spectra are sensitive to the scale height above a cloud deck at 0.1 - 1.0 bar. If the temperature above the cloud deck is substantially higher than the equilibrium temperature ($\sim$200 K) of the planet, the strength of the absorption features will be proportionally larger (Figure~\ref{JWST}). Such a stratosphere commonly exists in all giant planets in the Solar System, and has recently been detected in one hot exoplanet \citep{Evans2017}, although the degree of heating with respect to their equilibrium temperatures differs from planet to planet. Figure~\ref{JWST}b  shows  models with and without a heated stratosphere (HS) and two levels of metallicity, solar and 10 $\times$ Solar. The effect of stratospheric heating is much stronger than enhanced metallicity.

Figure~\ref{JWST}  shows that $JWST$ could detect the strong CH$_4$ features at 2.3 and 3.4 $\mu$m at low-to-moderate confidence in several models. We expect that spectral retrieval algorithms  \citep[e.g.,][]{LWZ13} will likely provide a higher confidence detection of CH$_4$ since such methods combine information on all features in the observed spectrum. We do not expect $HST$ observations to yield detections of CH$_4$ or other features in the model spectra. The smaller aperture of $HST$ will produce lower SNR in the $1.1 - 1.7\mu$m passband of its WFC3 G141 instrument mode than for $JWST$ NIRSpec. The transmission models show no spectral absorption features and only modest Raleigh slopes at wavelengths shorter than $\lambda = 600$ nm ($JWST$/NIRSpec's lower cutoff), so shorter wavelength HST observations will also not be able to constrain the planet's atmospheric properties.  The $JWST$ NIRSpec prism spectra could certainly detect the spectral features in the heated stratosphere models.

Finally, we put Kepler-1654b  into the context of other long period transiting systems suitable for observation by JWST. Table~\ref{LongP} gives data on  18 confirmed planets with radius $\geq 2$R$_\oplus$, orbital periods greater than 250  days and an equilibrium temperature\footnote{We adopt an illustrative equilibrium temperature given by $T_{pl}=265 L^{0.25} d^{-0.5}$ K for a planet located at d(AU) from a star of luminosity, $L(L_\odot)$} T$_{pl}<$300K. We developed a figure of merit which takes into account the total number of stellar photons, denoted $S$, observed in a spectral element, $\Delta\nu$, in a time $\tau$; the photon shot-noise, $\sigma=\sqrt{S}$; and the transit depth, $\alpha$. The ``Transit SNR" is defined as $\alpha S/ \sigma=\alpha\sqrt{S}$ and is evaluated for stellar flux densities, $F_\nu$, at K$_s$ (2.2 $\mu$m) or  WISE W2 (4.6 $\mu$m) for a telescope with a  collecting area $A$=25 m$^2$, with an instrument of resolution $R$=100 and efficiency $\eta$=0.25, and  in an integration time,  $\tau$, equal to the duration of a transit: $S=F_\nu A\eta\Delta\nu \tau /(h\nu)$. This figure of merit glosses over many details \citep{Greene2016}, but serves to rank these planets in terms of their suitability for transit spectroscopy. For planets with a temperature below 200 K,  only Kepler-167e, which is a larger planet orbiting a smaller star \citep{Kipping2016}, has a ``Transit SNR" larger than Kepler-1654b's. Other systems rank a factor of two or more lower, making Kepler-1654b a valuable target for future study. Of course, the atmospheric scale height  which depends on the planet's temperature and surface gravity  also affects the detectability of spectral signatures. But since only a few  of these planets have RV-determined masses  we do not account for the effects of  scale height here. 

The last column of Table~\ref{LongP} highlights the challenge of actually observing these long period planets. The long time between transits and JWST's  limited pointing  windows can make scheduling  difficult. JWST's sun avoidance restrictions determine when the $10^o\times10^o$  Kepler field can be observed,  nominally from early/mid-April to late-October/mid-November. Thus, for example,  transits of Kepler-1654b and Kepler 167e will be observable only starting with the 2024 events based on extrapolations from the information in the JWST APT tool.  

 \begin{deluxetable*}{ccccccccccc}
\tablecaption{ Properties of Long Period Transiting Planets \label{LongP}}
\tabletypesize{\scriptsize}
\tablehead{\colhead{Planet}&\colhead{Period}&\colhead{R$_{pl}$}&\colhead{Depth}&\colhead{Duration}&\colhead{Ks}&\colhead{WISE2}&\colhead{SNR$^*$}&\colhead{SNR$^*$}&\colhead{T$_{pl}$}&\colhead{First }\\
\colhead{Name}&\colhead{(days)} & \colhead{(R$_{Jup}$)}&\colhead{(ppm)}&\colhead{(days)}&\colhead{(mag)}&\colhead{(mag)}&\colhead{(Ks)}&\colhead{ (W2)}&\colhead{(K)}&\colhead{JWST}}
\startdata
Kepler-167e$^1$ & 1,070 & 0.91 & 16,224 & 0.67 & 11.83 & 11.84 & 407 & 213 & 140 & 2024-10-25\\
PH2b/Kepler 86b$^9$ & 280 & 0.90 & 8,589 & 0.44 & 11.12 & 11.14 & 242 & 126 & 284 & 2020-10-28\\
Kepler-553c$^6$ & 330 & 1.00 & 14,549 & 0.51 & 13.06 & 12.88 & 180 & 103 & 234 & 2019-06-08\\
{\bf Kepler-1654b}$^2$ & 1,410 & 0.82 & 5,095 & 0.89 & 11.92 & 11.93 & 141 & 74 & 177 & 2024-10-30$^{11}$\\
Kepler-421b$^4$ & 700 & 0.37 & 2,510 & 0.66 & 11.54 & 11.49 & 71 & 38 & 177 & 2025-10-10\\
Kepler-1647b$^3$ & 1,110 & 1.06 & 3,687 & 0.41 & 12.00 & 11.90 & 67 & 37 & 255 & 2021-08-02\\
Kepler-1625b$^6$ & 290 & 0.54 & 3,489 & 0.79 & 13.92 & 13.92$^{12}$ & 36 & 19 & 275 & 2019-05-26\\
KIC 9663113b$^5$ & 570 & 0.41 & 1,669 & 0.83 & 12.50 & 12.46 & 34 & 18 & 244 & 2020-10-23\\
Kepler-1536b$^6$ & 360 & 0.28 & 1,840 & 0.54 & 12.55 & 12.54 & 30 & 16 & 176 & 2019-05-12\\
KIC 10525077b$^5$ & 850 & 0.49 & 2,489 & 0.83 & 13.75 & 13.80 & 29 & 15 & 211 & 2019-04-11\\
Kepler-1630b$^6$ & 510 & 0.20 & 1,009 & 0.35 & 11.80 & 11.71 & 18 & 10 & 165 & 2019-07-15\\
Kepler-22b$^{10}$ & 290 & 0.21 & 493 & 0.31 & 10.15 & 10.15 & 18 & 10 & 272 & 2019-09-08\\
Kepler-1634b$^6$ & 370 & 0.29 & 1,080 & 0.47 & 12.72 & 12.68 & 15 & 8 & 238 & 2019-08-02\\
Kepler-150f$^7$ & 640 & 0.33 & 1,259 & 0.56 & 13.37 & 13.37 & 14 & 7 & 207 & 2024-05-09\\
Kepler-1635b$^6$ & 470 & 0.33 & 1,540 & 0.56 & 13.90 & 13.90 & 14 & 7 & 212 & 2020-06-12\\
Kepler-1600b$^6$ & 390 & 0.28 & 1,219 & 0.41 & 13.90 & 13.88 & 9 & 5 & 218 & 2019-10-06\\
Kepler-1632b$^6$ & 450 & 0.22 & 360 & 0.53 & 11.66 & 11.64 & 9 & 5 & 281 & 2020-05-15\\
Kepler-1636b$^6$ & 430 & 0.29 & 840 & 0.74 & 14.23 & 14.23$^{12}$ & 7 & 4 & 255 & 2023-05-23\\
\enddata
\tablecomments{Notes: $^*$See text for a description of the ``Transit SNR" figure of merit in R=100 spectral element.  $^1$\citet{Kipping2016}; $^2$This work; $^3$Circumbinary planet with multiple transits \citet{Kostov2016}; $^4$\citet{Kipping2014}; $^5$\citet{Wang2015}; $^6$\citet{Morton2016}; $^7$\citet{Schmitt2017}.  $^8$\citet{Jenkins2015}.  $^9$\citet{Wang2013}.  $^{10}$\citet{Borucki2012}.  $^{11}$This transit is just at the edge of the JWST observability window based on current knowledge. $^{12}$Estimated from 2MASS.}
\end{deluxetable*}

\section{Conclusion}

We have searched Q1-Q17 Kepler light curves of F and G  stars not previously associated with confirmed or candidate planets or even with Kepler "Objects of Interest" and  we were able to identify Kepler-1654b (originally KIC~8410697b)  which shows  two transits with a 1047 day period---one of the longest periods yet found in the Kepler survey. Subsequent AO and RV observations were able to rule out false positives and to characterize the planet and its host star. A fit to the combined transit curve plus RV data shows that orbiting this  mature G5 star is a 0.82 \rj\ planet with a mass of $<$0.5 M$_{Jup}$.   Transit spectroscopy  with JWST of Kepler-1654b and similar objects will enable a careful study of planets whose physical states, e.g. a low equilibrium temperature of $\sim$200 K, most closely resemble  those  of the outer planets in our own solar system.

%\clearpage

\section{Acknowledgements}
Some of the research described in this publication was carried out in part at the Jet Propulsion Laboratory, California Institute of Technology, under a contract with the National Aeronautics and Space Administration. This research has made use of the NASA/IPAC Infrared Science Archive (IRSA), the Keck Observatory Archive (KOA), and  the NASA Exoplanet Archive which are operated by the Jet Propulsion Laboratory, California Institute of Technology, under contract with the National Aeronautics and Space Administration.  We used the implementation of  EXOFAST available at the NASA Exoplanet Science Institute.

We are grateful to an anonymous referee for a careful reading of the manuscript which led to a number of improvements. Some data presented herein were obtained at the W. M. Keck Observatory from telescope time allocated to the National Aeronautics and Space Administration through the agency's scientific partnership with the California Institute of Technology and the University of California. The Observatory was made possible by the generous financial support of the W. M. Keck Foundation. The authors wish to recognize and acknowledge the very significant cultural role and reverence that the summit of Maunakea has always had within the indigenous Hawaiian community. We are most fortunate to have the opportunity to conduct observations from this mountain. Finally, HG acknowledges support of a  summer internship made possible by Caltech and JPL.

Copyright 2018 California Inst of Technology. All rights reserved.

\end{document}